\documentclass[12pt]{article}
\usepackage{amssymb}
\usepackage{amsmath}
\usepackage{epsf}
\usepackage{epsfig}
\usepackage{cite}
\newcommand{\lsim}{
\mathrel{\hbox{\rlap{\hbox{\lower4pt\hbox{$\sim$}}}\hbox{$<$}}}}

\newcommand{\gsim}{
\mathrel{\hbox{\rlap{\hbox{\lower4pt\hbox{$\sim$}}}\hbox{$>$}}}}

\def\D0{D\O }

\def\BsKK{B_s\to K^+ K^-}
\def\Amix{\mathcal{A}^{\text{mix}}_{\text{CP}}}
\def\Adir{\mathcal{A}^{\text{dir}}_{\text{CP}}}
\def\Adel{\mathcal{A}_{\Delta\Gamma}}

\begin{document}
\begin{titlepage}
\vspace*{-0.0truecm}

\begin{flushright}
Nikhef-2010-039
\end{flushright}

\vspace*{1.8truecm}

\begin{center}
\boldmath
{\Large{\bf In Pursuit of New Physics with $B^0_s\to K^+K^-$}}
\unboldmath
\end{center}

\vspace{0.9truecm}

\begin{center}
{\bf Robert Fleischer  \,and\, Robert Knegjens}

\vspace{0.5truecm}

{\sl Nikhef, Science Park 105, NL-1098 XG Amsterdam, The Netherlands}
 
\end{center}

\vspace{1.6cm}
\begin{abstract}
\vspace{0.2cm}\noindent
The $B^0_s\to K^+K^-$ decay is the $U$-spin partner of $B^0_d\to\pi^+\pi^-$ and allows a 
determination of the angle $\gamma$ of the unitarity triangle of the Cabibbo--Kobayashi--Maskawa 
matrix. Using updated information on the branching ratio and the relevant hadronic form factors, 
we discuss the picture that emerges for $\gamma$ from the currently available data, which is in 
good agreement with the fits of the unitarity triangle. We point out that the 
$B^0_s\to K^+K^-$ decay also offers interesting probes to search for new physics in 
$B^0_s$--$\bar B^0_s$ mixing, thereby complementing the well-known $B^0_s\to J/\psi \phi$ 
analyses. The relevant observables are the effective lifetime of this channel and its 
mixing-induced CP asymmetry. We calculate correlations between these quantities and the 
CP-violating $B^0_s$--$\bar B^0_s$ mixing phase, which serve as target regions for improved measurements at the Tevatron and the early data taking at the LHCb experiment. Finally, we 
discuss the expected situation for the optimal determination of $\gamma$ at LHCb, exploiting
precise measurements of the CP violation in $B^0_s\to K^+K^-$.
\end{abstract}

\vspace*{0.5truecm}
\vfill
\noindent
November 2010
\vspace*{0.5truecm}

\end{titlepage}

\thispagestyle{empty}
\vbox{}
\newpage

\setcounter{page}{1}

\section{Introduction}\label{sec:intro}
The decay $B^0_s\to K^+K^-$, which is governed by QCD penguins and has a
doubly Cabibbo-suppressed tree contribution in the Standard Model (SM), 
has a very interesting physics potential for $B$-decay experiments at hadron colliders. 
It is related to the $B^0_d\to\pi^+\pi^-$ channel through the $U$-spin flavour symmetry of 
strong interactions, which allows a determination of the angle $\gamma$ of the unitarity 
triangle of the Cabibbo--Kobayashi--Maskawa (CKM) matrix and certain hadronic parameters 
\cite{RF-BsKK}. The advantage of this $U$-spin strategy with respect to the conventional
$SU(3)$ flavour-symmetry strategies is twofold: 
\begin{itemize}
\item no additional dynamical assumptions
have to be made (such as the neglect of particular topologies), which could be spoiled
by large rescattering effects;
\item electroweak penguins, which are not invariant under the isospin symmetry because
of the different up- and down-quark charges, are automatically included.
\end{itemize}
In Ref.~\cite{RF-Bhh}, a detailed discussion of the status of the extraction of $\gamma$ 
coming from the first measurement of the $B^0_s\to K^+K^-$ branching ratio at the 
Tevatron was given, addressing also tests of the $U$-spin symmetry.

In the current paper, we have a fresh look at the determination of $\gamma$ both in view of 
updated information on the $B^0_s\to K^+K^-$ branching ratio and in view of updated information 
on the relevant $U$-spin-breaking form-factor ratio. 
The main focus of this paper, however, is to point out the usefulness of measurements of the $B^0_s\to K^+K^-$ channel's effective lifetime $\tau_{K^+K^-}$ and mixing-induced CP asymmetry for probing New Physics (NP) in $B^0_s$--$\bar B^0_s$ mixing. 
This analysis provides target ranges for these observables 
for improved measurements at the Tevatron and the LHCb experiment at CERN's Large Hadron 
Collider (LHC) and complements nicely analyses of CP-violating effects in the angular distributions
of the decay products of $B^0_s\to J/\psi [\to \mu^+\mu^-]\phi[\to K^+K^-]$ \cite{DDF,DFN}. 
With respect to an analysis of various $B^0_s\to K^+K^-$ correlations performed almost
a decade ago \cite{FlMa}, we now obtain a significantly sharper picture. 

The CDF and \D0 collaborations have used $B^0_s\to J/\psi \phi$ measurements to obtain the first direct results for  the CP-violating $B^0_s$--$\bar B^0_s$ mixing phase $\phi_s$.
In the SM this quantity is fixed in terms of the Wolfenstein parameters \cite{wolf} by $-2\lambda^2\eta\sim -2^\circ$ but could be enhanced significantly through NP effects. 
Unfortunately, the most recent data are still far 
from being conclusive: while 
the CDF collaboration finds $\phi_s\in [-59.6^\circ,-2.29^\circ]\sim-30^\circ \, \lor \, 
[-177.6^\circ,-123.8^\circ]\sim-150^\circ$ (68\% C.L.) \cite{CDF-phis}, \D0 quotes a best 
fit value around $\phi_s\sim -45^\circ$ \cite{D0-phis}, taking also information form the 
dimuon charge asymmetry and the measured $B_s\to D_s^{(*)+}D_s^{(*)-}$ branching 
ratio into account. As we will show in the current paper, the $B^0_s\to K^+K^-$ channel 
offers an excellent alternative probe for such effects, thereby nicely complementing the 
$B^0_s\to J/\psi \phi$ analysis and allowing us also to resolve a twofold discrete
ambiguity for the extracted value of $\phi_s$. 

The outline of this paper is as follows: in Section~\ref{sec:gam}, we discuss the picture for 
$\gamma$ arising from the current data. In Section~\ref{sec:lifetime}, we calculate the effective 
lifetime $\tau_{K^+K^-}$ and study its correlation with the CP-violating $B^0_s$--$\bar B^0_s$ 
mixing phase. A similar analysis is performed for the mixing-induced CP asymmetry of 
$B^0_s\to K^+K^-$ in Section~\ref{sec:ACPmix}. In Section~\ref{sec:gam-optimal}, we 
discuss and illustrate the optimal determination of $\gamma$ using the CP asymmetries 
of $B^0_s\to K^+K^-$. Finally, we summarize our conclusions in Section~\ref{sec:concl}.

\boldmath
\section{Determination of $\gamma$}\label{sec:gam}
\unboldmath
In the SM, the $B_s^0\to K^+K^-$ and $B^0_d\to\pi^+\pi^-$ decay amplitudes can be 
written as follows \cite{RF-BsKK}:
\begin{equation}\label{BsKK-ampl}
A(B_s^0\to K^+K^-)=e^{i\gamma}\lambda\,{\cal C}'\left[1+\frac{1}{\epsilon}
d'e^{i\theta'}e^{-i\gamma}\right]
\end{equation}
\begin{equation}\label{Bdpipi-ampl}
A(B_d^0\to\pi^+\pi^-)=e^{i\gamma}\left(1-\frac{\lambda^2}{2}\right){\cal C}
\left[1-d\,e^{i\theta}e^{-i\gamma}\right],
\end{equation}
where $\lambda\equiv|V_{us}|= 0.22543\pm0.00077$ is the Wolfenstein parameter
\cite{CKMfitter}, 
$\epsilon\equiv\lambda^2/(1-\lambda^2)$, ${\cal C}$ and ${\cal C'}$ are 
CP-conserving strong amplitudes that are governed by the tree contributions, 
while the CP-consering hadronic parameters $d e^{i\theta}$ and $d' e^{i\theta'}$
measure -- loosely speaking -- the ratio of penguin to tree amplitudes.

If we apply the $U$-spin symmetry, we obtain the relations \cite{RF-BsKK}
\begin{equation}\label{U-spin-rel-1}
d'=d, \quad \theta'=\theta.
\end{equation}
As was pointed out in Ref.~\cite{RF-BsKK}, these relations are not affected
by factorizable $U$-spin-breaking corrections, i.e.\ the relevant form factors and
decay constants cancel. This feature holds also for chirally enhanced contributions
to the transition amplitudes. 

On the other hand, the $U$-spin symmetry also implies $|{\cal C}'/{\cal C}|=1$. Here, however, the corresponding decay constants 
and form factors do not cancel, so that we obtain the following result in the
factorization approximation:
\begin{equation}\label{QCDSR}
\left|\frac{{\cal C}'}{{\cal C}}\right|_{\rm fact}=
\frac{f_K}{f_\pi}\frac{F_{B_sK}(M_K^2;0^+)}{F_{B_d\pi}(M_\pi^2;0^+)}
\left(\frac{M_{B_s}^2-M_K^2}{M_{B_d}^2-M_\pi^2}\right) .
\end{equation}
Using the updated QCD light-cone sum rule calculation of Ref.~\cite{DuMe} yields
\begin{equation}
\left|\frac{{\cal C}'}{{\cal C}}\right|^{\rm QCDSR}_{\rm fact}=1.41^{+0.20}_{-0.11},
\end{equation}
which is consistent within the errors with the numerical value of $1.52^{+0.18}_{-0.14}$
obtained previously by the authors of Ref.~\cite{KKM} that was used in Ref.~\cite{RF-Bhh}.

The $B^0_s\to K^+K^-$ decay is now well established and the 
Heavy Flavour Averaging Group (HFAG) gives the following average for its 
branching ratio \cite{HFAG}:
\begin{equation}\label{BsKK-HFAG}
\mbox{BR}(B^0_s\to K^+K^-)=(26.5\pm4.4)\times10^{-6},
\end{equation}
which is a combination of measurements by CDF \cite{CDF-BsKK} at the Tevatron 
and by Belle \cite{Belle} at KEKB runs at the $\Upsilon(5S)$ resonance. A major source
of uncertainty for the branching ratio measurement of any $B_s$ decay is the ratio 
$f_s/f_d$ of fragmentation functions $f_q$ describing the probability that a $b$ quark will 
hadronize as a $\bar B_q$ meson. Using a newly proposed strategy \cite{FST}, this ratio 
can be measured at LHCb with an expected uncertainty that is two times smaller
than that of current compilations in the literature. 

For the extraction of $\gamma$, it is useful to introduce the following ratio of CP-averaged 
branching ratios:
\begin{eqnarray}
K&=&\frac{1}{\epsilon}\,\left|\frac{{\cal C}}{{\cal C}'}\right|^2
\left[\frac{M_{B_s}}{M_{B_d}}\,\frac{\Phi(M_\pi/M_{B_d},M_\pi/M_{B_d})}{
\Phi(M_K/M_{B_s},M_K/M_{B_s})}\,\frac{\tau_{B_d}}{\tau_{B_s}}\right]
\left[\frac{\mbox{BR}(B_s\to K^+K^-)}{\mbox{BR}(B_d\to\pi^+\pi^-)}\right]\nonumber\\
&=&\frac{1}{\epsilon^2}\left[\frac{\epsilon^2+2\epsilon d'\cos\theta'\cos\gamma+
d'^2}{1-2d\cos\theta\cos\gamma+d^2}\right]\stackrel{\rm exp}{=}51.8^{+12.7}_{-15.0},\label{K-def}
\end{eqnarray}
where we have used (\ref{BsKK-ampl}) and (\ref{Bdpipi-ampl}). The numerical value 
results from a combination of (\ref{QCDSR}) and (\ref{BsKK-HFAG}) with the HFAG 
average $\mbox{BR}(B_d\to\pi^+\pi^-)=(5.16\pm0.22)\times 10^{-6}$,  where we 
have added all errors in quadrature. 

The $B^0_s\to K^+K^-$ and $B^0_d\to \pi^+\pi^-$ decays are into CP-even eigenstates 
and offer the following time-dependent CP asymmetries:
\begin{eqnarray}
\lefteqn{\frac{\Gamma(B^0_q(t)\to f)-
\Gamma(\bar B^0_q(t)\to f)}{\Gamma(B^0_q(t)\to f)+
\Gamma(\bar B^0_q(t)\to f)}}\nonumber\\
&&=\left[\frac{{\cal A}_{\rm CP}^{\rm dir}(B_q\to f)\,\cos(\Delta M_q t)+
{\cal A}_{\rm CP}^{\rm mix}(B_q\to f)\,\sin(\Delta 
M_q t)}{\cosh(\Delta\Gamma_qt/2)+{\cal A}_{\rm 
\Delta\Gamma}(B_q\to f)\,\sinh(\Delta\Gamma_qt/2)}\right],\label{ACP-t}
\end{eqnarray}
where $\Delta M_q\equiv M_{\rm H}^{(q)} -M_{\rm L}^{(q)}$ and 
$\Delta\Gamma_q\equiv\Gamma_{\rm L}^{(q)}-\Gamma_{\rm H}^{(q)}$
are the mass and width differences of the 
``heavy" and ``light" $B_q$ mass eigenstates, respectively. The width difference is
negligible in the $B_d$ system but is expected at the $15\%$ level in the $B_s$ case 
within the SM, as we will discuss in more detail in Section~\ref{sec:lifetime}. Note that for 
the sign definition of $\Delta\Gamma_q$ given above, the Standard-Model value
$\Delta\Gamma_s^{\rm SM}$ is positive.

In the case of the $B^0_s\to K^+K^-$ channel, we unfortunately do not yet have 
a measurement of the CP asymmetry in (\ref{ACP-t}) available. On the other hand, 
measurements of the CP-violating observables of the  $B^0_d\to\pi^+\pi^-$ channel 
have been performed at the $B$ factories. By using (\ref{Bdpipi-ampl}), we can derive the expressions 
\begin{eqnarray}
{\cal A}_{\rm CP}^{\rm dir}(B_d\to \pi^+\pi^-)&=&
-\left[\frac{2\,d\sin\theta\sin\gamma}{1-
2\,d\cos\theta\cos\gamma+d^2}\right],\nonumber\\
{\cal A}_{\rm CP}^{\rm mix}(B_d\to \pi^+\pi^-)&=&+\left[\,
\frac{\sin(\phi_d+2\gamma)-2\,d\,\cos\theta\,\sin(\phi_d+\gamma)+
d^2\sin\phi_d}{1-2\,d\cos\theta\cos\gamma+d^2}\,\right],\label{ACP-Bpipi}
\end{eqnarray}
where $\phi_d$ denotes the CP-violating $B^0_d$--$\bar B^0_d$ mixing phase which is 
measured by means of CP violation in $B_d\to J/\psi K_{\rm S,L}$ type decays. 
Taking also hadronic corrections from doubly Cabibbo-suppressed penguin contributions through $B^0_d\to J/\psi\pi^0$ data into account gives \cite{FFJM}
\begin{equation}\label{eqn:phiD}
\phi_d=(42.4^{+3.4}_{-1.7})^\circ.
\end{equation}

The current experimental status of the CP violation in $B^0_d\to\pi^+\pi^-$ can be summarized
as 
\begin{equation}
{\cal A}_{\rm CP}^{\rm mix}(B_d\to\pi^+\pi^-)=
\left\{\begin{array}{ll}
0.68\pm0.10\pm0.03 & \mbox{(BaBar \cite{babar-pipi})}\\
0.61\pm0.10\pm0.04 & \mbox{(Belle \cite{belle-pipi}),}
\end{array}\right.
\end{equation}
\begin{equation}\label{ACP-dir}
{\cal A}_{\rm CP}^{\rm dir}(B_d\to\pi^+\pi^-)=\left\{
\begin{array}{cc}
-0.25\pm0.08\pm0.02 & \mbox{(BaBar \cite{babar-pipi})}\\
-0.55\pm0.08\pm0.05 & \mbox{(Belle \cite{belle-pipi}).}
\end{array}\right.
\end{equation}
While there is a good agreement between BaBar and Belle for the measurement of
the mixing-induced CP asymmetry, yielding the average \cite{HFAG}
\begin{equation}
{\cal A}_{\rm CP}^{\rm mix}(B_d\to\pi^+\pi^-)=0.65\pm0.07,
\end{equation}
we are unfortunately still facing a discrepancy between the BaBar and Belle results for
the direct CP asymmetry, which, in our opinion, makes it problematic to combine them 
in an average. This feature is reflected by the averages appearing in the literature: 
HFAG gives $-0.38 \pm 0.06$, whereas the Particle Data Group (PDG) quote the same 
central value with a larger error of 0.17 \cite{PDG}.

\begin{figure}[t]%  figure placement: here, top, bottom, or page
   \centering
   \begin{tabular}{cc}
   	  \includegraphics[width=7.4truecm]{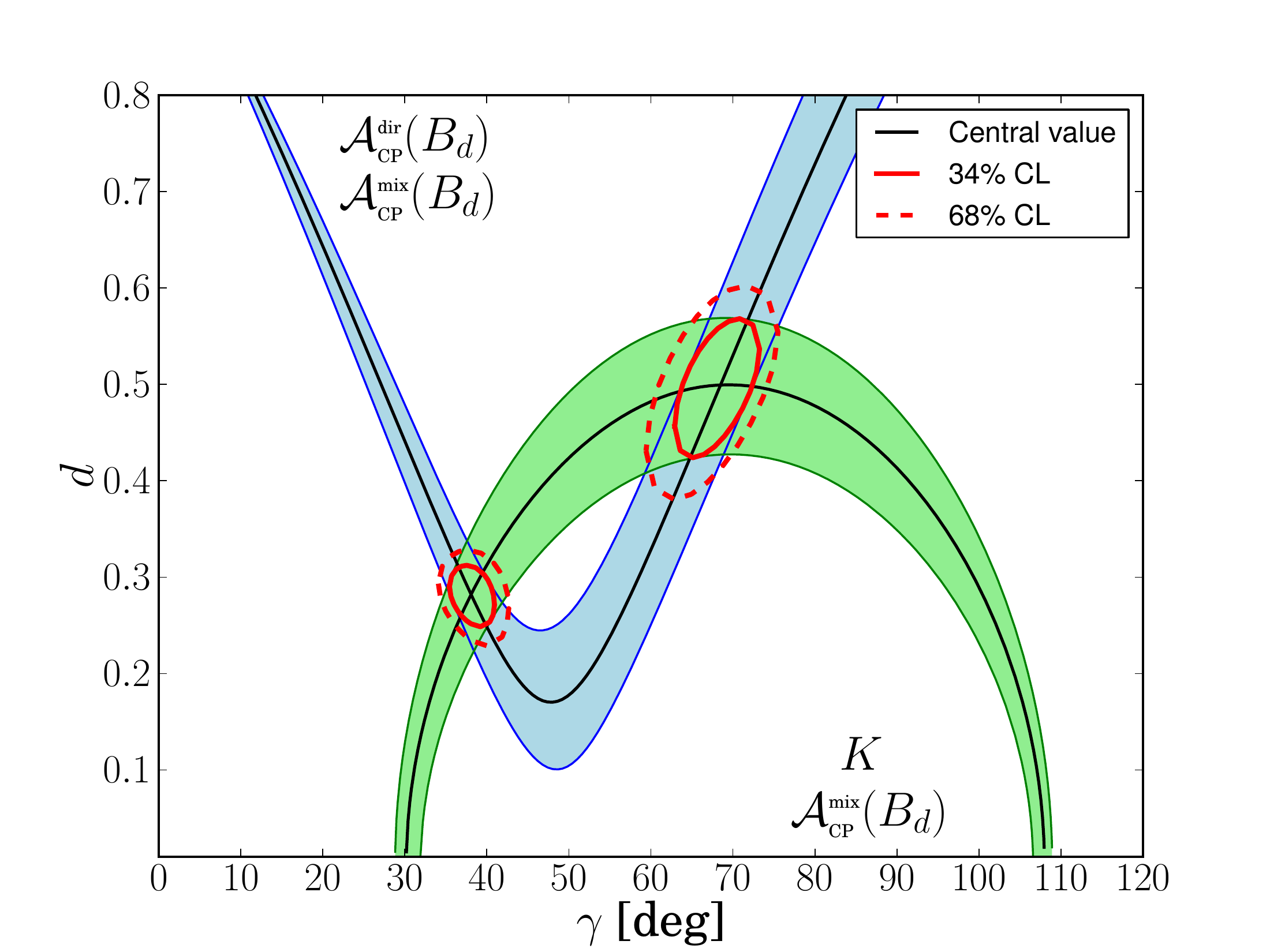} &
      \includegraphics[width=7.4truecm]{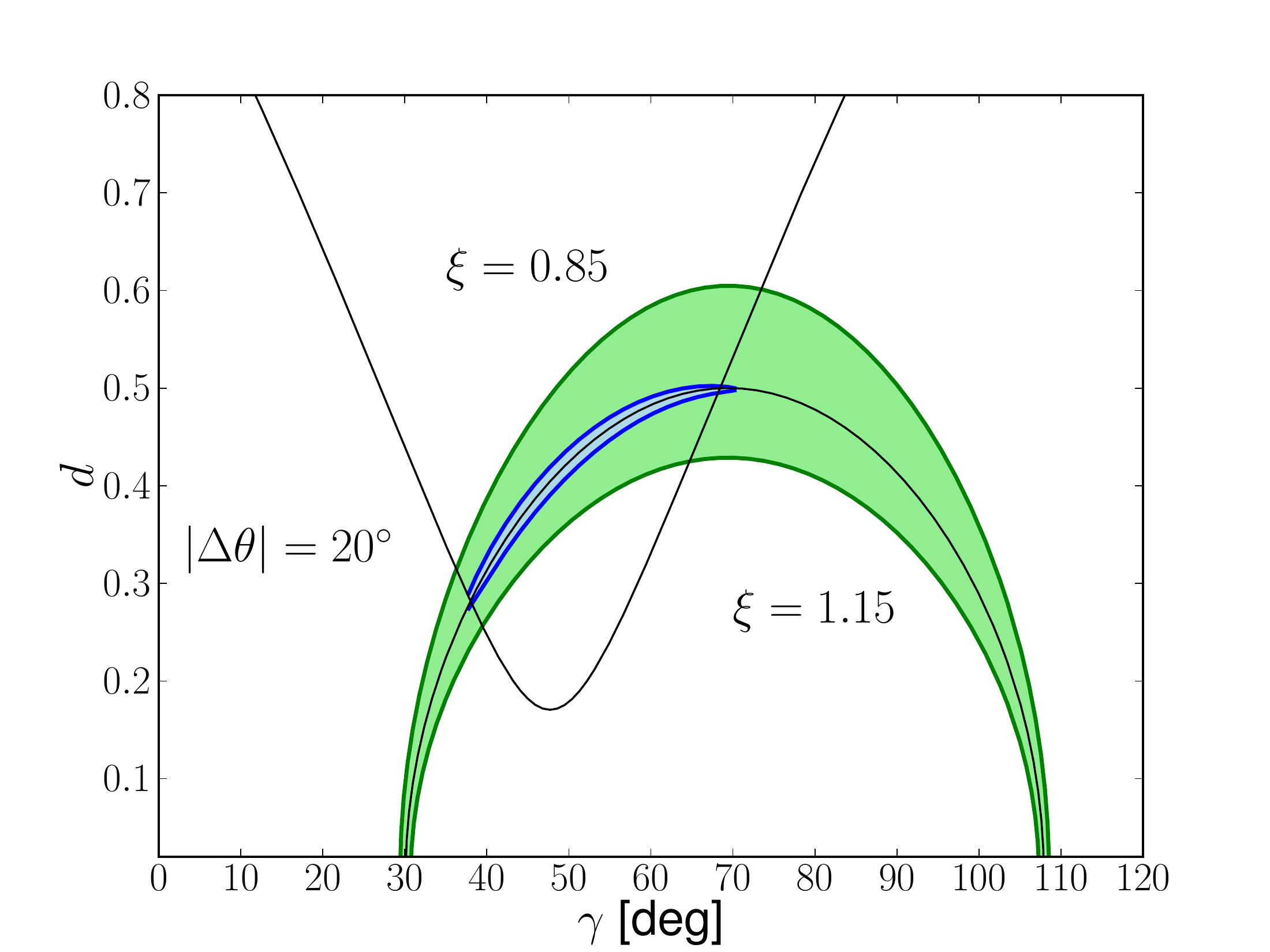} 
     \end{tabular}
   \caption{The contours in the $\gamma$--$d$ plane fixed through the CP-violating
    $B^0_d\to\pi^+\pi^-$ observables and $K$. Left panel: $1\,\sigma$
error bands and the 68\% C.L. regions, right panel: illustration of $U$-spin-breaking effects.}
   \label{fig:1}
\end{figure}

In view of this unsatisfactory situation, we would like to avoid these averages. An
alternative is offered by the direct CP violation in $B^0_d\to \pi^-K^+$, 
where all measurements are consistent with one another, yielding the average 
${\cal A}_{\rm CP}^{\rm dir}(B_d\to\pi^\mp K^\pm)=0.098^{+0.011}_{-0.012}$
(for a sign convention consistent with (\ref{ACP-t})) \cite{HFAG}. The $B^0_d\to \pi^-K^+$  and 
$B^0_s\to K^+K^-$ channels differ only in their spectator quarks. 
We can neglect exchange and penguin annihilation topologies, as these are expected to be tiny and can further be constrained by the absence of anomalous
behaviour in $B^0_d\to K^+K^-$ and $B^0_s\to \pi^+\pi^-$ data \cite{RF-Bhh}.
Consequently, $SU(3)$ flavour symmetry implies the following relation:
\begin{eqnarray}
\lefteqn{{\cal A}_{\rm CP}^{\rm dir}(B_d\to\pi^+\pi^-)=}\nonumber\\
&&-\left(\frac{f_\pi}{f_K}\right)^2
\left[\frac{\mbox{BR}(B_d\to\pi^\mp K^\pm)}{\mbox{BR}(B_d\to\pi^+\pi^-)}\right]
{\cal A}_{\rm CP}^{\rm dir}(B_d\to\pi^\mp K^\pm)=-0.26\pm0.03\label{Adir-rel1},
\end{eqnarray}
which is in excellent agreement with the BaBar result (\ref{ACP-dir}) (see 
also Ref.~\cite{FRS}). 

In the proceeding numerical analysis, we shall use 
${\cal A}_{\rm CP}^{\rm dir}(B_d\to\pi^+\pi^-)=-0.26\pm0.10$, i.e.\ we increase the error
of (\ref{Adir-rel1}) generously to allow for possible $SU(3)$-breaking corrections. 
It will turn out that this CP asymmetry plays a minor role for the error budget of the extraction of $\gamma$.

\begin{figure}[t]%  figure placement: here, top, bottom, or page
   \centering
     \includegraphics[width=12.8truecm]{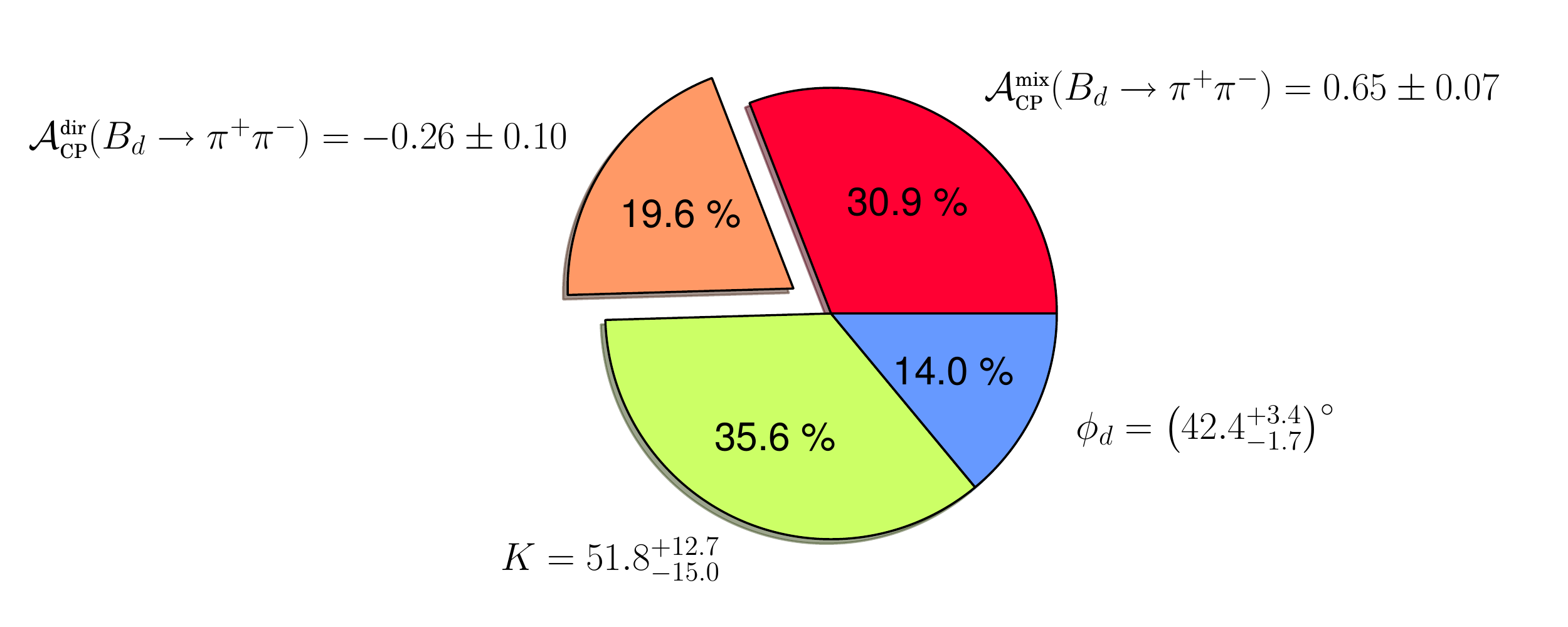} 
   \caption{The error budget for the extracted value of $\gamma$ in (\ref{gam-extr}) that 
   is associated with the used input data.} \label{fig:2}
\end{figure}

In order to determine $\gamma$, we can convert the direct and mixing-induced CP asymmetry
of the $B^0_d\to\pi^+\pi^-$ channel into a theoretically clean contour in the
$\gamma$--$d$ plane; the corresponding formulae are given in Ref.~\cite{RF-BsKK}.
Furthermore, using the $U$-spin relation (\ref{U-spin-rel-1}) in (\ref{K-def}), we can determine a second
contour. The intersection of both contours then allows us to determine $\gamma$ and $d$, 
so that we can also extract the strong phase $\theta$. In Fig.~\ref{fig:1}, we show the
situation arising for the current data. 
The plot on the left-hand side shows the $1\,\sigma$ error bands and the 34\% and 68\% confidence regions arising from a $\chi^2$ fit, whereas the plot on the right-hand side illustrates the impact of $U$-spin-breaking corrections to (\ref{U-spin-rel-1}), which we have parameterized as 
\begin{align}
\xi\equiv d'/d &= 1 \pm 0.15, \label{U-break:xi} \\
\Delta\theta\equiv\theta'-\theta &= \pm 20^\circ. \label{U-break:theta}
\end{align}
As discussed in Ref.~\cite{RF-Bhh}, the discrete ambiguity can be resolved, yielding 
\begin{equation}\label{gam-extr}
\gamma=(68.3^{+4.9}_{-5.7}|_{\rm input}
	\mbox{}^{+5.0}_{-3.7}|_\xi\mbox{}^{+0.1}_{-0.2}|_{\Delta\theta})^\circ
\end{equation}
and
\begin{equation}\label{eqn:gam-hadr}
 	d = 0.499^{+0.069}_{-0.076}|_{\rm input}
	\mbox{}^{+0.101}_{-0.074}|_\xi\mbox{}^{+0.002}_{-0.005}|_{\Delta\theta},
	\, \quad
	\theta =  (153.7^{+10.8}_{-13.6}|_{\rm input}
	\mbox{}^{+3.8}_{-3.9}|_\xi\mbox{}^{+0.1}_{-0.2}|_{\Delta\theta})^\circ,
\end{equation}
where the input errors are the 68\% confidence intervals of the $\chi^2$ fit.
In Fig.~\ref{fig:2}, we show the error budget for $\gamma$ coming from the 
individual input quantities. 
We observe that $K$ and ${\cal A}_{\rm CP}^{\rm mix}(B_d\to\pi^+\pi^-)$ have a similar impact on the error, while ${\cal A}_{\rm CP}^{\rm dir}(B_d\to\pi^+\pi^-)$  plays a significantly less important role.
This is a nice feature in view of the unsatisfactory experimental situation corresponding to the direct measurement of this observable discussed above. 
Using the central value from the HFAG and PDG averages of the BaBar and Belle data, 
 ${\cal A}_{\rm CP}^{\rm dir}(B_d\to\pi^+\pi^-)=-0.38$, yields a central value of
 $\gamma=65^\circ$, which is fully consistent with (\ref{gam-extr}). Interestingly, also the error of $\phi_d$ given in \eqref{eqn:phiD} has a small but non-negligible impact on the overall error. In Ref.~\cite{BFK}, a strategy to include penguin effects in the determinations of $\phi_d$ using $B^0_s\to J/\psi K_{\rm S}$ was discussed, allowing us to match the experimental precision for the measurement of that phase at LHCb. 
 
The extracted value given in (\ref{gam-extr}) is in excellent agreement with the fits 
of the unitarity triangle, yielding
\begin{equation}\label{gamma-fit}
\gamma=
\left\{\begin{array}{ll}
(67.2^{+3.9}_{-3.9})^\circ & \mbox{(CKMfitter \cite{CKMfitter})}\\
(69.6\pm3.1)^\circ & \mbox{(UTfit \cite{UTfit}).}
\end{array}\right.
\end{equation}
In view of this feature, large NP effects at the amplitude level are already excluded
by the current data. We shall therefore assume the SM expressions in 
(\ref{BsKK-ampl}) and (\ref{Bdpipi-ampl}) for the following discussion.

The determination of $\gamma$ can be significantly improved once a measurement of
CP-violating observables of the $B^0_s\to K^+K^-$ channel are available. We will 
discuss this in more detail in Section~\ref{sec:gam-optimal}. Let us next have a 
closer look at the effective $B^0_s\to K^+K^-$ lifetime, which is particularly interesting
for improved measurements by CDF at the Tevatron and the early data taking at LHCb.

\boldmath
\section{The Effective $B^0_s\to K^+K^-$ Lifetime}\label{sec:lifetime}
\unboldmath
The ``untagged" rate of initially, i.e.\ at time $t=0$, present $B^0_s$ or $\bar B^0_s$ 
decays into the $K^+K^-$ final state can be written as follows \cite{DF}:
\begin{align}
	\langle \Gamma(B_s(t)\to K^+K^-)\rangle
	\equiv&\ \Gamma(B^0_s(t)\to K^+K^-)+ \Gamma(\bar B^0_s(t)\to K^+K^-) 
	\notag\\
	=&\ R_{\rm H}(B_s\to K^+K^-)e^{-\Gamma_{\rm H}^{(s)} t} +  R_{\rm L}(B_s\to K^+K^-)
	e^{-\Gamma_{\rm L}^{(s)} t}\label{untagged}\\
	\propto &\ e^{-\Gamma_st}\left[ \cosh\left(\frac{\Delta\Gamma_s t}{2}\right)+
	{\cal A}_{\rm \Delta\Gamma}(B_s\to K^+K^-)\,\sinh\left(\frac{\Delta\Gamma_s t}
	{2}\right)\right],\notag
\end{align}
where 
\begin{equation}
\Gamma_s\equiv \frac{\Gamma_{\rm H}^{(s)}+\Gamma_{\rm L}^{(s)}}{2}=\tau_{B_s}^{-1}, 
\end{equation}
with $\tau_{B_s}$ denoting the $B^0_s$ lifetime.  The observable 
${\cal A}_{\Delta\Gamma}(B_s\to K^+K^-)$, which enters also the 
time-dependent CP asymmetry (\ref{ACP-t}), is given by
\begin{equation}
{\cal A}_{\Delta\Gamma}(B_s\to K^+K^-)=
\frac{R_{\rm H}(B_s\to K^+K^-)-R_{\rm L}(B_s\to K^+K^-)}{R_{\rm H}(B_s\to K^+K^-)+
R_{\rm L}(B_s\to K^+K^-)}.
\end{equation}
Using the parametrization in (\ref{BsKK-ampl}), we obtain
\begin{equation}\label{ADG-BsKK}
{\cal A}_{\Delta\Gamma}(B_s\to K^+K^-)=
-\left[\frac{d'^2\cos\phi_s+2\epsilon d'\cos\theta'\cos(\phi_s+\gamma)
+\epsilon^2\cos(\phi_s+2\gamma)}{d'^2+2\epsilon d'\cos\theta'\cos\gamma+
\epsilon^2}\right].
\end{equation}

A particularly nice and simple observable that is offered by the $B^0_s\to K^+K^-$ decay 
is its effective lifetime, which is defined as 
\begin{equation}
 \tau_{K^+ K^-} \equiv \frac{\int^\infty_0 t\ \langle \Gamma(B_s(t)\to K^+K^-)\rangle\ dt}
  {\int^\infty_0 \langle \Gamma(B_s(t)\to K^+K^-)\rangle\ dt}.
\end{equation}
This quantity is also the resulting lifetime if the untagged rate with the two exponentials
in (\ref{untagged}) is fitted to a single exponential \cite{DFN}. Using the two-exponential
form in (\ref{untagged}) with $ R_{\rm H,L}\equiv  R_{\rm H,L}(B_s\to K^+K^-)$ yields 
\begin{equation}
  \tau_{K^+ K^-} =\frac{ R_{\rm L}/\Gamma_{\rm L}^{(s)2}+ R_{\rm H}/\Gamma_{\rm H}^{(s)2}}
  {R_{\rm L}/\Gamma_{\rm L}^{(s)}+ R_{\rm H}/\Gamma_{\rm H}^{(s)}},
\end{equation}
which can be written in terms of 
\begin{equation}
y_s\equiv\frac{\Delta\Gamma_s}{2\Gamma_s},
\end{equation}
the observable $\Adel\equiv\Adel(B_s\to K^+K^-)$, and the $B^0_s$ lifetime as
\begin{equation}	\label{eqn:lifetime}
	\frac{\tau_{K^+K^-}}{\tau_{B_s}}
	=\frac{1}{1-y_s^2} \left[\frac{1+2\Adel y_s+y_s^2}{1+ \Adel y_s}\right] 
	=\ 1 + \Adel y_s
	+\left(2- \Adel^2\right)y_s^2
	+ {\cal O}\left(y_s^3\right).
\end{equation}
First studies along these lines were performed by the CDF collaboration in 2006 
\cite{CDF-DGBsKK}, yielding a value of 
\begin{equation}\label{tau-exp}
\tau_{K^+K^-}= (1.53 \pm 0.18 \pm 0.02)\,\mbox{ps}.
\end{equation}
Unfortunately, this analysis has, to the best of our knowledge, not been updated, which we strongly encourage in view of the results discussed below.

In the following analysis, we assume that $\gamma$ is known, with a value of $\gamma=(68\pm7)^\circ$ in agreement with \eqref{gam-extr} and the fits of the UT in (\ref{gamma-fit}). 
Using then $K$ as given in \eqref{K-def} with the $U$-spin relation \eqref{U-spin-rel-1} 
and the direct CP asymmetry
\begin{equation}
{\cal A}_{\rm CP}^{\rm dir}(B_s\to K^+K^-) = 
\frac{2\epsilon d'\sin\theta'\sin\gamma}{d'^2+
2\epsilon d'\cos\theta'\cos\gamma+\epsilon^2},
\end{equation}
we can determine $d'$ and $\cos\theta'$, allowing us to calculate ${\cal A}_{\Delta\Gamma}(B_s\to K^+K^-)$ for a given value of the $B^0_s$--$\bar B^0_s$ mixing phase $\phi_s$. 
Parametrizing the $U$-spin-breaking effects for $d^{(')}$ through \eqref{U-break:xi}, the corresponding formulae from which $d'$ and $\cos\theta'$ can be extracted are 
\begin{align}
	2d'\cos\theta' 
	=&\ \frac{-d'^2 + \epsilon^2\left[K(1+d'^2\xi^{-2})-1\right]}
	{\epsilon\cos\gamma\left(1+ \epsilon  \xi^{-1}K\right)}, \\
	d'^2 =&\ \epsilon^2\left[\frac{b \pm \sqrt{b^2-ac}}{a}\right],
\end{align}
with
\begin{align}
 	a =&\ \epsilon^2\xi^{-2}\left(1+\epsilon\xi^{-1}\right)^2 (\Adir)^2 K^2\cot^2\gamma  
	+ \left(1-\epsilon^2\xi^{-2} K \right)^2, \\
	b =&\ -\epsilon\xi^{-1} \left(1+\epsilon\xi^{-1}\right)^2 (\Adir)^2 K^2\cot^2\gamma  
	+2\cos^2 \gamma \left(1+\epsilon\xi^{-1}K\right)^2 \\
	&\ +(K-1)(1- \epsilon^2\xi^{-2} K ),\\
	c =&\  \left(1+\epsilon\xi^{-1}\right)^2 (\Adir)^2 K^2 \cot^2\gamma + (K-1)^2, 
	\label{eqn:d2Amix}
\end{align}
where $\Adir\equiv\Adir(B_s\to K^+ K^-) $.
The $U$-spin-breaking effects of $\theta^{(')}$, as given in \eqref{U-break:theta}, are difficult to include in the above analytic expressions but are straightforward to calculate numerically.

As we have noted above, the CP-violating observables of the $B^0_s\to K^+K^-$ channel have not yet been measured.
However, as the $B^0_d\to \pi^-K^+$ and $B^0_s\to K^+K^-$ channels differ only in their spectator quarks (see the discussion of (\ref{Adir-rel1})), we expect
\begin{equation}\label{Adir-SM}
{\cal A}_{\rm CP}^{\rm dir}(B_s\to K^+K^-)\approx 
{\cal A}_{\rm CP}^{\rm dir}(B_d\to\pi^\mp K^\pm)=0.098^{+0.011}_{-0.012}.
\end{equation}
In order to take possible corrections into account, we increase the error generously and use 
\begin{equation}\label{ABsKK-input}
{\cal A}_{\rm CP}^{\rm dir}(B_s\to K^+K^-)=0.098\pm0.04
\end{equation}
in the following numerical analysis.
 
Given all of these inputs, including the $U$-spin breaking effects \eqref{U-break:xi} and \eqref{U-break:theta}, we arrive at the following results for the hadronic parameters:
\begin{equation}
  	d'  	  = 0.499^{+0.085}_{-0.101}, \quad
	\cos\theta'	  = -0.886^{+0.131}_{-0.089},
\end{equation}
where all errors were added in quadrature.

\begin{figure}[t]%  figure placement: here, top, bottom, or page
   \centering
   \begin{tabular}{cc}
   	  \includegraphics[width=7.4truecm]{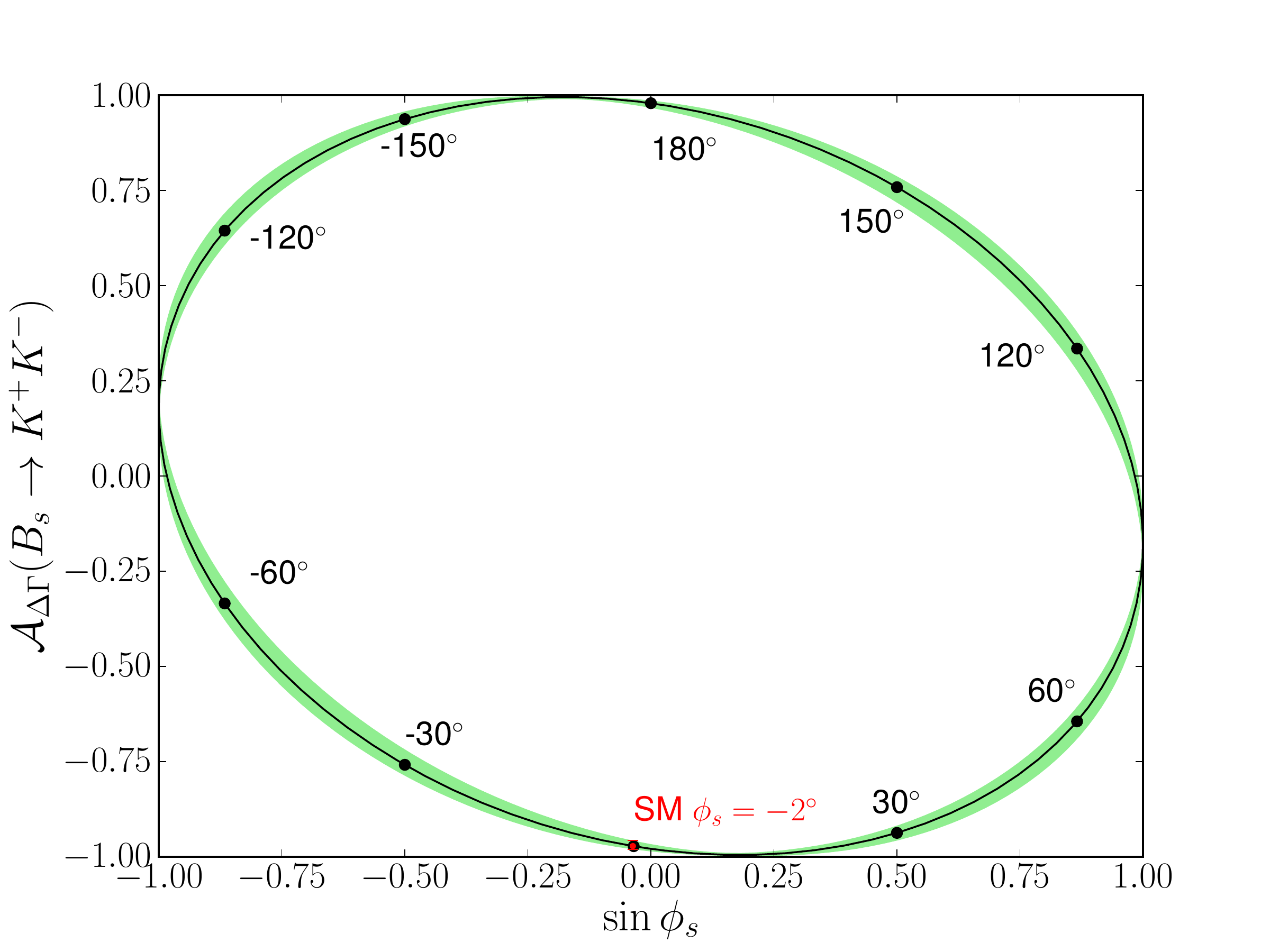} &
      \includegraphics[width=7.4truecm]{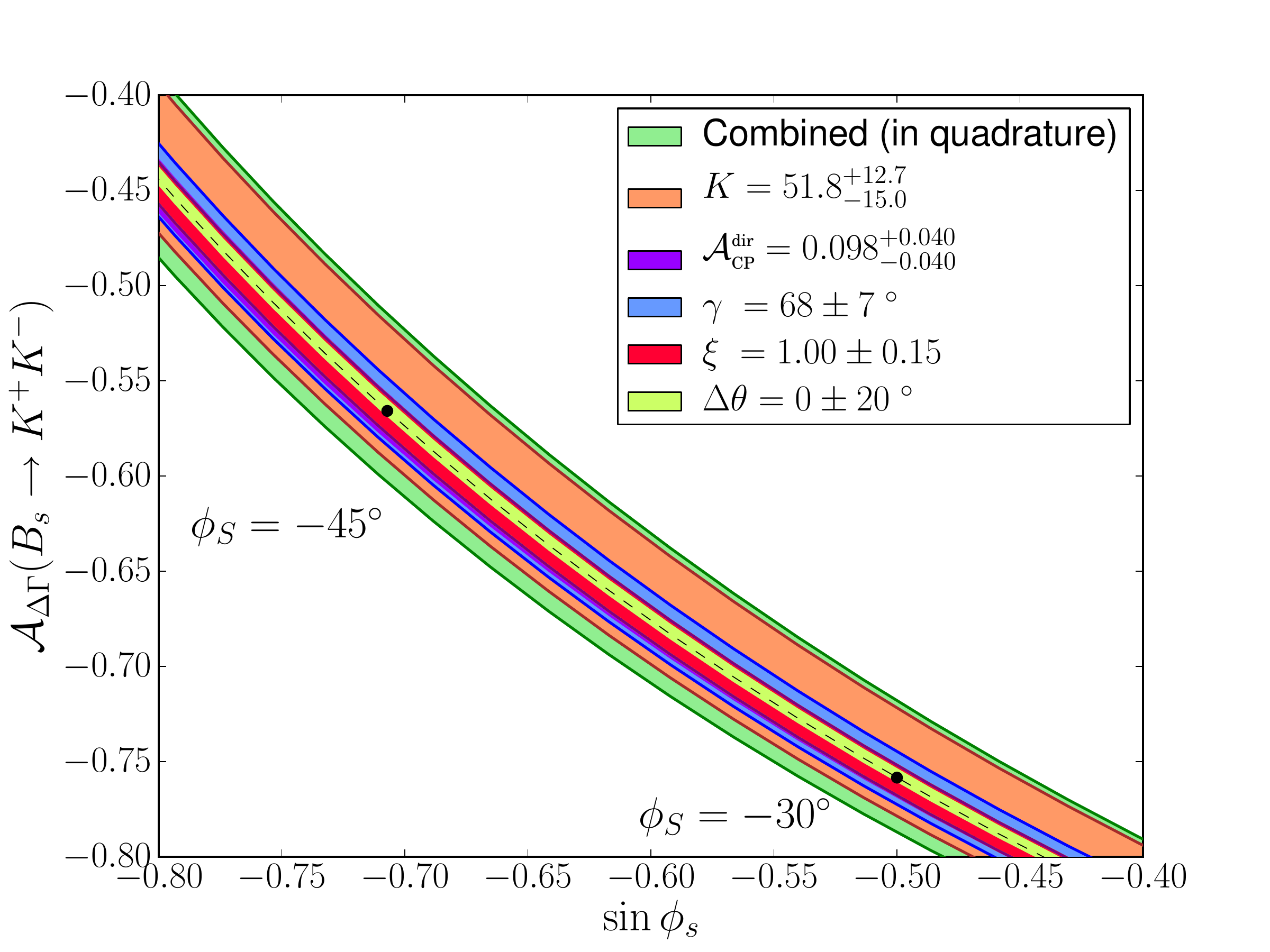} 
     \end{tabular}
   \caption{Left panel: correlation between ${\cal A}_{\Delta\Gamma}(B_s\to K^+K^-)$ and $\sin\phi_s$. Right panel: errors associated with the input observables/parameters, zoomed in for a NP region of $\phi_s\in[-25^\circ,-50^\circ]$ and overlayed on top of one another. The legend lists the error contributions from largest to smallest. }\label{fig:ADG}
\end{figure}

The SM prediction for the $B^0_s$--$\bar B^0_s$ mixing phase is $\phi_s=-(2.1\pm0.1)^\circ$
\cite{CKMfitter,UTfit}, where the error has an essentially negligible impact on the following 
analysis. The corresponding  prediction for the ``untagged" observable is
\begin{align}\label{ADG-SM}
	{\cal A}_{\Delta\Gamma}(B_s\to K^+K^-)\Big|_{\rm SM}
	=&\ -0.97178
		{}^{+0.0133}_{-0.0064}\Big|_{K} 
		{}^{+0.0047}_{-0.0046}\Big|_{\gamma}
		{}^{+0.00005}_{-0.00003}\Big|_{\Adir}
		{}^{+0.0022}_{-0.0031}\Big|_{\xi}
		{}^{+0.0016}_{-0.0008}\Big|_{\Delta\theta} \notag\\
	=&\ -0.972^{+0.014}_{-0.009},
\end{align}
where all errors have again been combined in quadrature.
Particularly interesting is the small influence of the $U$-spin breaking errors and ${\cal A}_{\rm CP}^{\rm dir}(B_s\to K^+K^-)$ on the total error. 
In Fig.~\ref{fig:ADG} we treat $\phi_s$ as a free parameter and show the correlation between 
${\cal A}_{\Delta\Gamma}(B_s\to K^+K^-)$ and $\sin\phi_s$, as well as errors related to the 
input quantities overlayed on top of one  another and centred on the central value. It is remarkable 
that ${\cal A}_{\Delta\Gamma}(B_s\to K^+K^-)$ is very robust with respect to the input errors
for the whole range of $\phi_s$.

\begin{figure}[t]%  figure placement: here, top, bottom, or page
   \centering
   \begin{tabular}{cc}
      \includegraphics[width=7.4truecm]{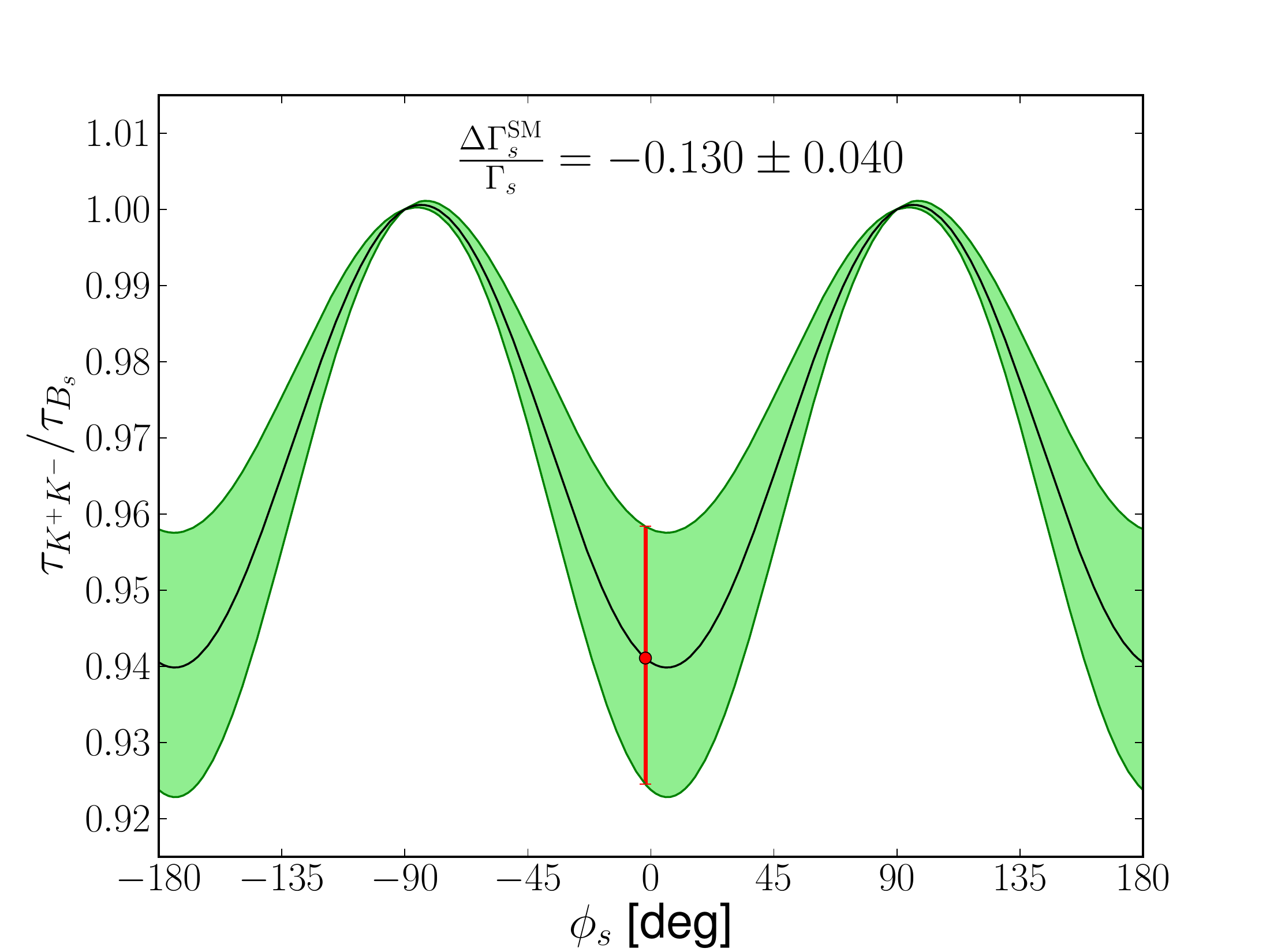} & 
   	  \includegraphics[width=7.4truecm]{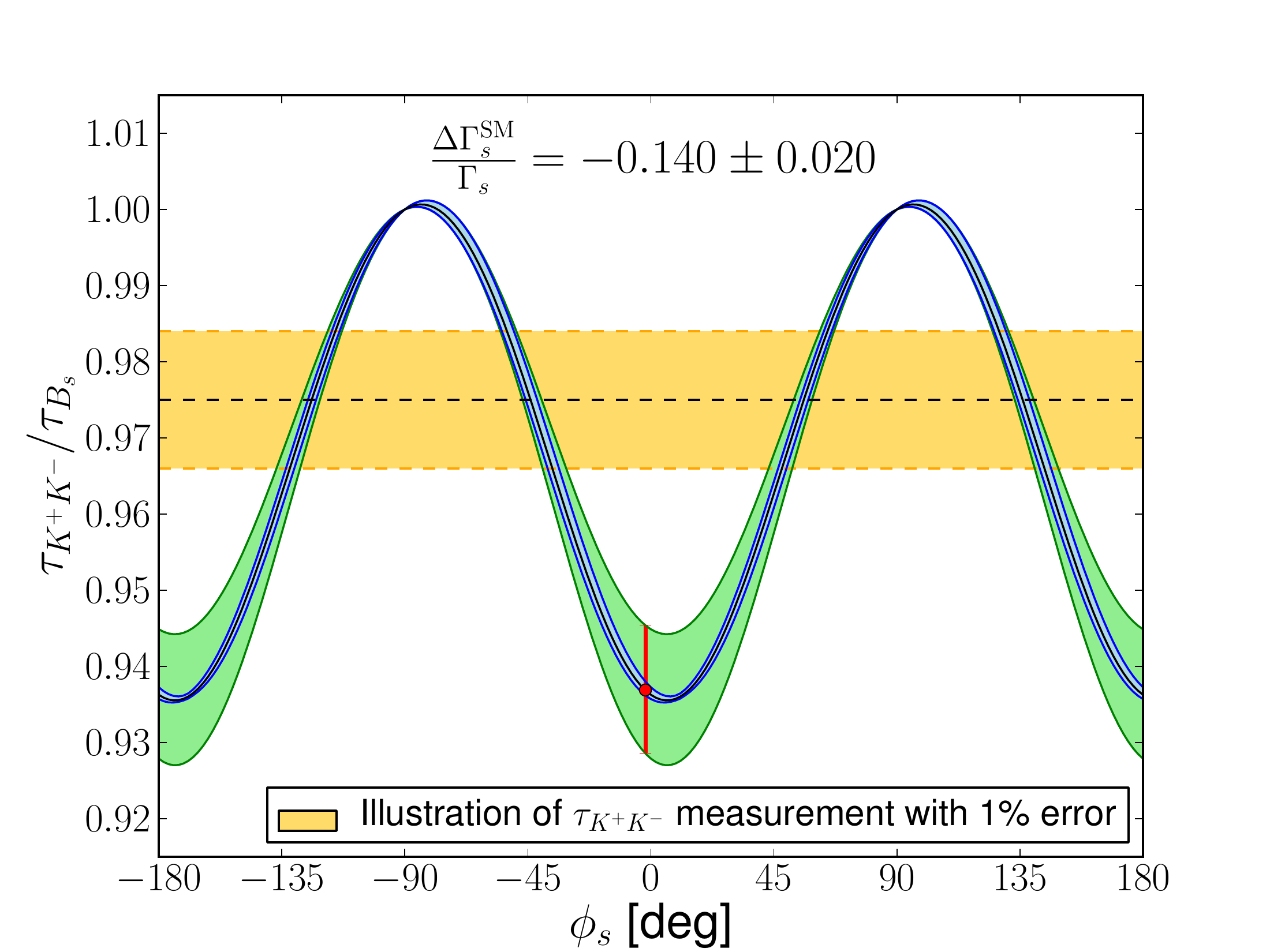} 
	\end{tabular}
   \caption{The dependence of $\tau_{K^+K^-}/\tau_{B_s}$ on the $B^0_s$--$\bar B^0_s$
   mixing phase. Left panel: uses $\Delta\Gamma_s^{\rm SM}/\Gamma_s$ with the larger 
   error in (\ref{DG-SM}), as in (\ref{ratio-SM}). Right panel: illustration of the impact of a 
   measurement of $\tau_{K^+K^-}/\tau_{B_s}$ with $1\%$ uncertainty (horizontal band) 
   for $\Delta\Gamma_s^{\rm SM}/\Gamma_s$ with the smaller error in (\ref{ratio-SM}). 
   The narrow band on the curve corresponds to the errors of the input quantities entering 
   our prediction of ${\cal A}_{\Delta\Gamma}(B_s\to K^+K^-)$.} 
   \label{fig:lifetime}
\end{figure}

The remaining ingredient for a determination of the effective lifetime is the width difference 
$\Delta\Gamma_s$. In the presence of CP-violating NP contributions to $B^0_s$--$\bar B^0_s$
mixing, it takes the following form \cite{Grossman:1996era}:
\begin{equation}\label{DG-phis}
	\Delta\Gamma_s
	= \Delta\Gamma_s^{\rm SM} \cos\phi_s,
\end{equation}
where $\Delta\Gamma_s^{\rm SM}$ is the width difference of the $B_s$-meson system
in the SM. The most recent updates for the theoretical results for this quantity are
\begin{equation}\label{DG-SM}
\frac{\Delta\Gamma_s^{\rm SM}}{\Gamma_s}=\left\{
\begin{array}{l}
0.13\pm0.04 \mbox{\cite{Nierste},} \\
0.14\pm 0.02 \mbox{\cite{Silvestrini}}.
\end{array}
\right.
\end{equation}
Using the expression in (\ref{eqn:lifetime}) with (\ref{ADG-SM}) and the value with the larger
error in (\ref{DG-SM}), we obtain the following SM prediction of the effective lifetime ratio:
\begin{equation}\label{ratio-SM}
  	\frac{\tau_{K^+K^-}}{\tau_{B_s}}\Big|_{\rm SM} = 0.9411 
		{}^{+0.0011}_{-0.0006}\Big|_{\Adel}
		{}^{+0.0173}_{-0.0165}\Big|_{\Delta\Gamma_s/\Gamma_s} \\
	= 0.941^{+0.017}_{-0.017},
\end{equation}
where the errors have been added in quadrature.
Combining this with the measurement $\tau_{B_s}=(1.477\pm0.022)$\,ps \cite{HFAG} gives a SM prediction for the lifetime of 
\begin{equation}
	\tau_{K^+K^-}|_{\rm SM} =(1.390\pm 0.032)\, {\rm ps},
\end{equation}
which is fully consistent with the CDF result in (\ref{tau-exp}) although the errors are too large
to draw any further conclusions at this point.
LHCb is expected to achieve a precision of $2\%$ (or better) with the data sample foreseen to be 
accumulated in 2011, corresponding to $1\mbox{fb}^{-1}$ at the LHC
centre-of-mass energy of 7 TeV \cite{Gersabeck}. 

In Fig.~\ref{fig:lifetime}, we show the dependence of $\tau_{K^+K^-}/\tau_{B_s}$ 
on the mixing phase $\phi_s$  and illustrate the impact of a measurement of this
lifetime ratio with a precision of $1\%$. Here we assume a value of $\phi_s=-45^\circ$, 
which is the best fit value of the current \D0 data (see Section~\ref{sec:intro}). 
The figure clearly shows that a measurement of the effective $B^0_s\to K^+K^-$ lifetime 
will offer an interesting alternative tool for finding evidence of a sizeable NP value for $\phi_s$, 
thereby complementing the ``conventional" analyses. The error band of the theoretical 
prediction is essentially fully dominated by that of $\Delta\Gamma_s^{\rm SM}/\Gamma_s$.
The discrete ambiguity for $\phi_s$ emerging from the lifetime analysis can be resolved
with the help of the mixing-induced CP violation in $B^0_s\to K^+K^-$, as we will see
in the next section.

As an alternative for using the theoretical SM value in (\ref{eqn:lifetime}), we may also use 
the experimental value of $\Delta\Gamma_s/\Gamma_s$, which depends implicitly 
on $\phi_s$ through (\ref{DG-phis}). By the end of 2011, LHCb is expected to reach
a sensitivity of about 0.03 for this measurement from $B^0_s\to J/\psi \phi$ \cite{Tristan}.
For a compilation of current experimental information of $\Delta\Gamma_s/\Gamma_s$,
assuming sometimes the SM, the reader is referred to Ref.~\cite{HFAG}.

\begin{figure}[t]%  figure placement: here, top, bottom, or page
   \centering
   \includegraphics[width=7.8truecm]{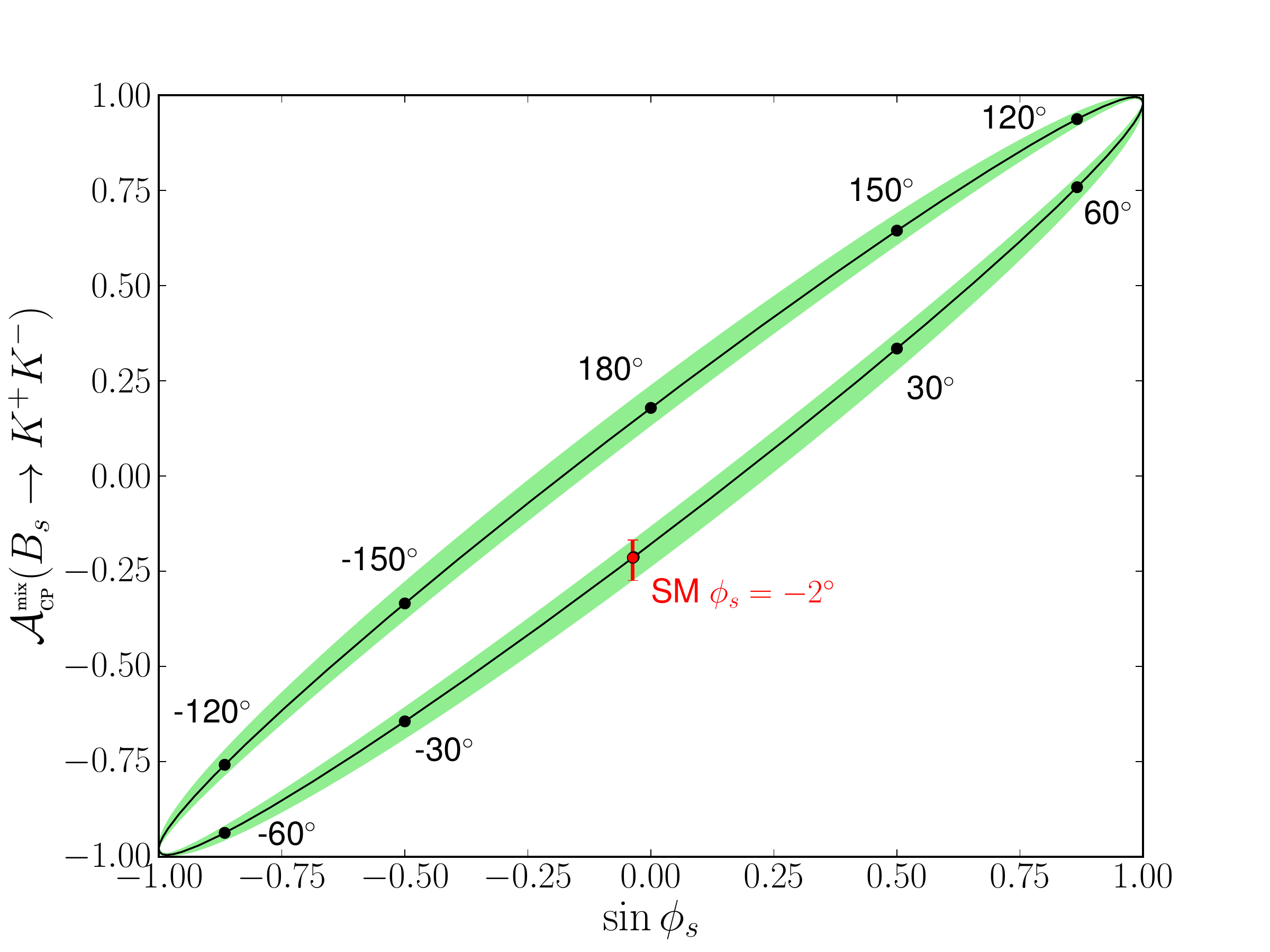} 
   \caption{The correlation between $\Amix(\BsKK)$ and $\sin\phi_s$.  
   The error band takes the uncertainties due to the input parameters
   and observables into account, as well as possible $U$-spin-breaking corrections. 
   The numbers label the values of $\phi_s$. }\label{fig:mixing}
\end{figure}

\boldmath
\section{Mixing-Induced CP Violation in $B^0_s\to K^+K^-$}\label{sec:ACPmix}
\unboldmath

\begin{figure}[t]%  figure placement: here, top, bottom, or page
   \centering
   \begin{tabular}{cc}
   	  \includegraphics[width=7.4truecm]{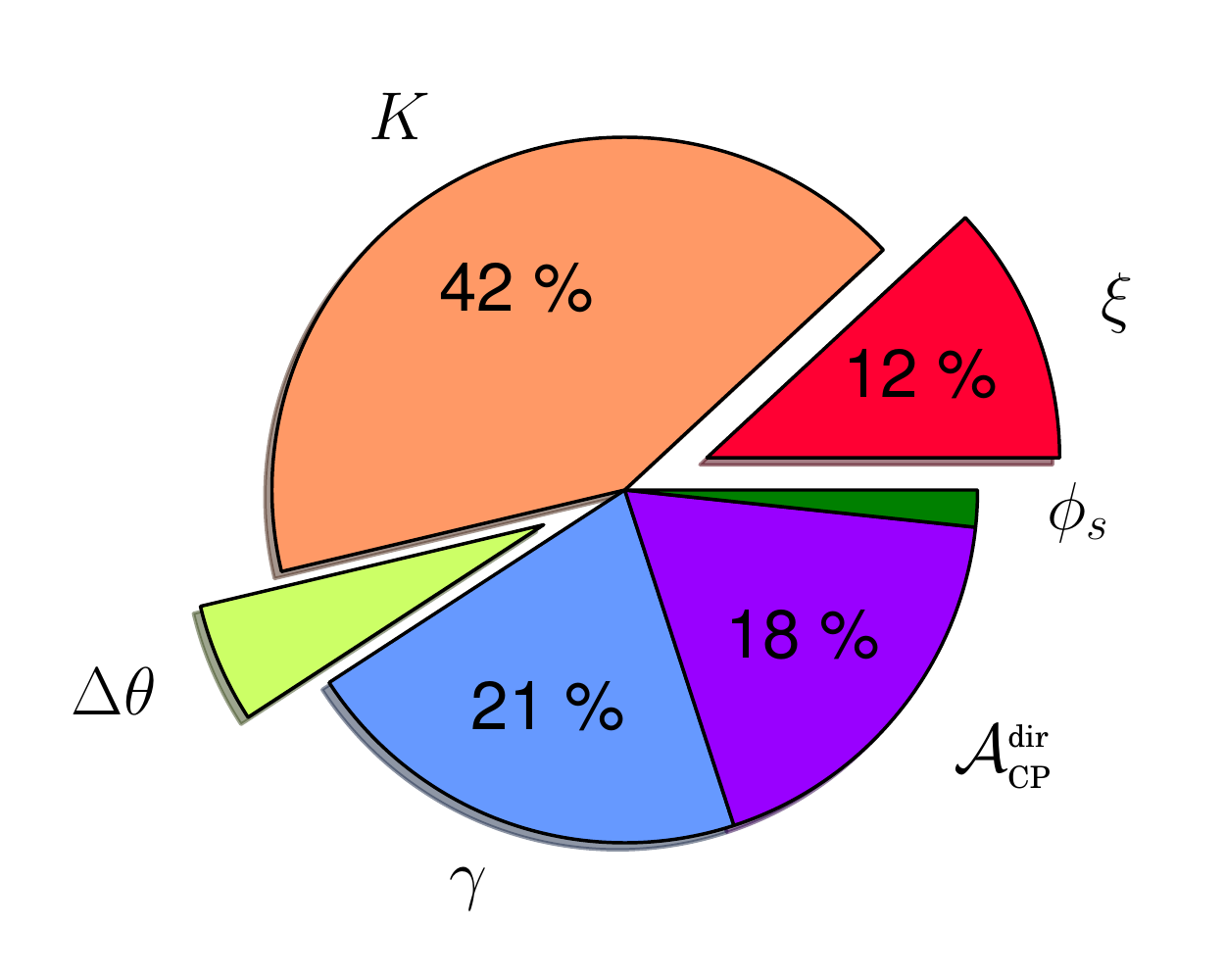} &
      \includegraphics[width=7.4truecm]{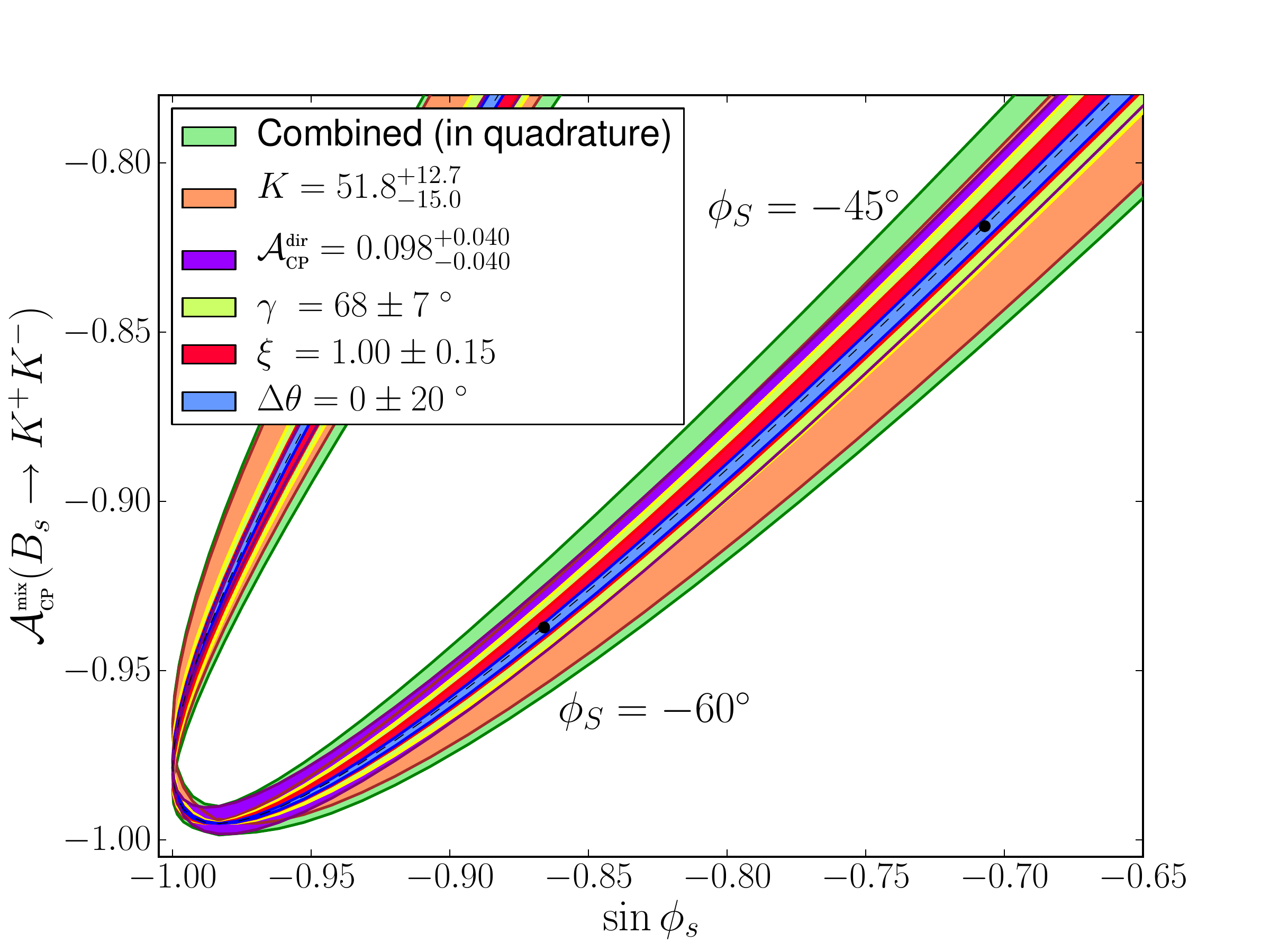} 
     \end{tabular}
   \caption{The error budget for $\Amix(\BsKK)$. Left panel: pie chart of the relative contribution
   of each input error for the SM case of $\phi_s^{\rm SM}$. Right panel: errors overlayed on top 
   of one another for the correlation in Fig.~\ref{fig:mixing}. The legend lists the error contributions from largest to smallest.}
   \label{fig:mixerrors}
\end{figure}

The final observable that is offered by the $B^0_s\to K^+K^-$ channel is its
mixing-induced CP asymmetry, which takes the following form \cite{RF-BsKK}:
\begin{equation}\label{Amix-BsKK}
{\cal A}_{\rm CP}^{\rm mix}(B_s\to K^+K^-)=+\left[\frac{d'^2\sin\phi_s + 
2\epsilon d' \cos\theta' \sin(\phi_s+\gamma)+ \epsilon^2\sin(\phi_s+
2\gamma)}{d'^2+2\epsilon d'\cos\theta'\cos\gamma+\epsilon^2}\right].
\end{equation}
The structure of this expression is very similar to (\ref{ADG-BsKK}), i.e.\ the strong
phase enters only as $\cos\theta'$. Consequently, we can use the formulae given 
in the previous section to perform an analysis of ${\cal A}_{\rm CP}^{\rm mix}(B_s\to K^+K^-)$
that is analogous to that of ${\cal A}_{\Delta\Gamma}(B_s\to K^+K^-)$. For the SM 
we obtain the prediction
\begin{align}
	\Amix(\BsKK)\Big|_{\rm SM} 
	=&\ -0.215 
		{}^{+0.031}_{-0.053}\Big|_{K} 
		{}^{+0.022}_{-0.020}\Big|_{\gamma}
		{}^{+0.023}_{-0.014}\Big|_{\Adir}
		{}^{+0.014}_{-0.010}\Big|_{\xi}
		{}^{+0.004}_{-0.007}\Big|_{\Delta\theta} \notag\\ 
	=&\ -0.215^{+0.047}_{-0.060},\label{Amix-SM}
\end{align}
where the errors have been combined in quadrature. In Fig.~\ref{fig:mixing}, we show the dependence of $\Amix(\BsKK)$ on $\sin\phi_s$, with a range of $\phi_s$ points marked 
explicitly. The latter quantity is conventionally measured through the CP-violating effects 
in the $B^0_s\to J/\psi\phi$ angular distribution, as discussed above. The error bars on the 
SM point correspond to those given above. This plot illustrates two interesting features:
\begin{itemize}
\item $\Amix(\BsKK)$ offers a powerful tool to search for footprints of a sizable NP phase
$\phi_s$, and deviates already significantly from the SM value for moderate values of this
phase (such as $\phi_s\sim-30^\circ$).
\item $\Amix(\BsKK)$ allows us to resolve the twofold ambiguity for the value of $\phi_s$ 
resulting from the analyses of $B^0_s\to J/\psi\phi$. In particular, we can then also distinguish
between the SM case with $\phi_s\sim 0^\circ$ and a NP scenario with $\phi_s\sim 180^\circ$,
both leading to small CP violation in $B^0_s\to J/\psi\phi$.
\end{itemize}
The correlation in Fig.~\ref{fig:mixing} was first discussed in Ref.~\cite{RF-Bhh}
(its counterpart for $B^0_s\to J/\psi K_{\rm S}$ was recently studied in Ref.~\cite{BFK}). 
Here we go beyond that analysis by making a detailed analysis of the 
corresponding errors and using $\gamma$ as an input. As in the previous section, 
we observe that the calculation is remarkably stable with respect to possible 
$U$-spin-breaking corrections and input errors. In Fig.~\ref{fig:mixerrors}, we show the 
error budget corresponding to the various input parameters and observables: for the 
SM case, we give a pie chart of the relative contribution of each error, and for a NP region, 
zoomed in on the range $\phi_s\in [-40^\circ,-120^\circ]$, we show, as in the right panel
of Fig.~\ref{fig:ADG}, the errors overlayed on top of one another and centred on the central value.

The experimental sensitivities of the CP-violating $B^0_s\to K^+K^-$ observables at LHCb were studied in Ref.~\cite{LHCbRoadMap}.
Already with an integrated luminosity of $200\,\mbox{pb}^{-1}$ at the 7 TeV run of the LHC, a statistical sensitivity for $\Amix(\BsKK)$ of 0.11 can be obtained at this experiment \cite{Wilkinson}. 
Consequently, already for the first LHCb measurement of CP violation in $B^0_s\to K^+K^-$ it will be exciting to confront the correlation in Fig.~\ref{fig:mixing} with real data.

\boldmath
\section{Optimal Determination of $\gamma$}\label{sec:gam-optimal}
\unboldmath

\begin{figure}[t]%  figure placement: here, top, bottom, or page
   \centering
   \includegraphics[width=7.8truecm]{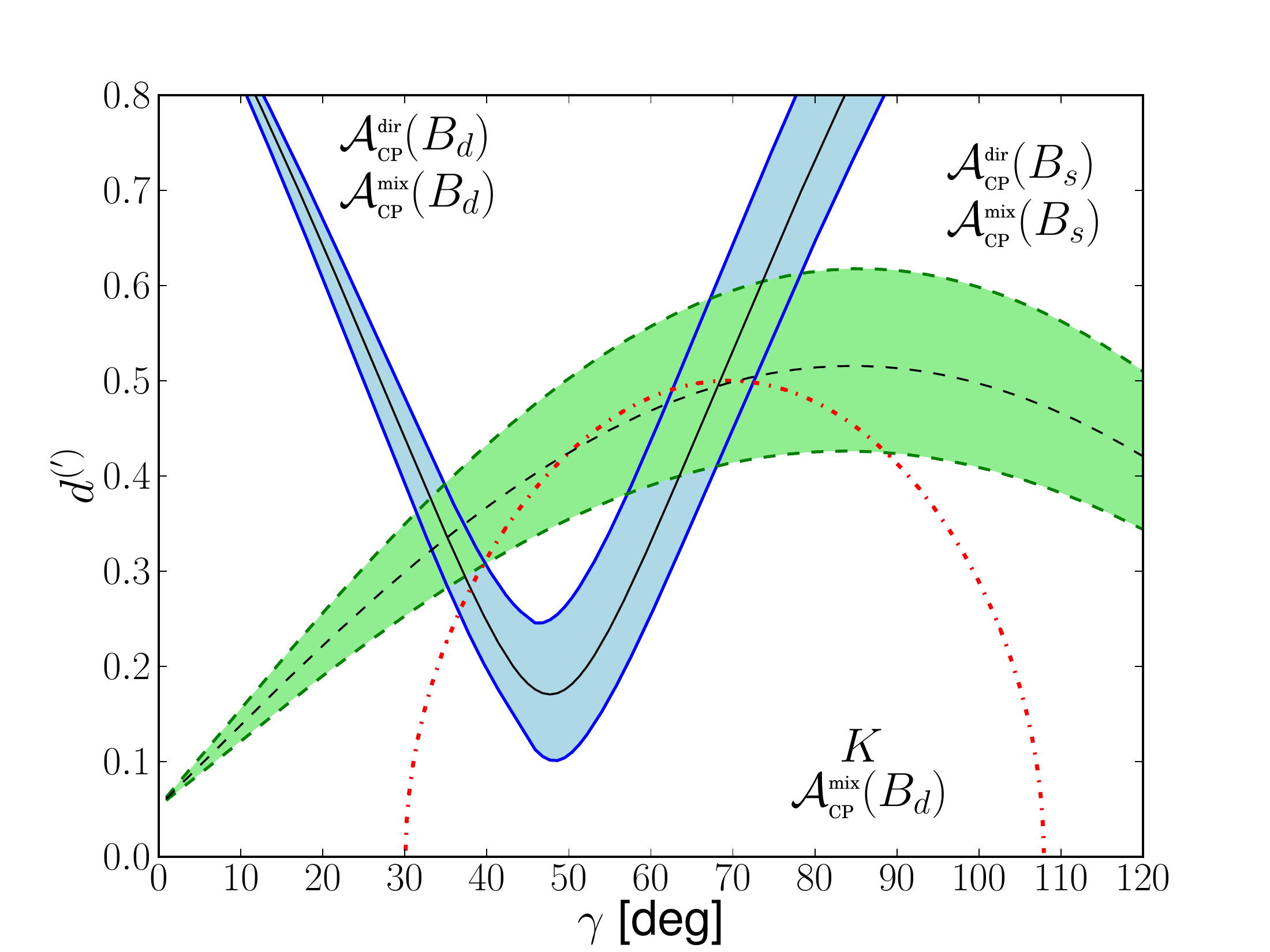} 
   \caption{Illustration of the optimal determination of $\gamma$ from the CP-violating observables of the
   $B_s\to K^+K^-$, $B_d\to\pi^+\pi^-$ system for the SM case. The band of the
   $B_s\to K^+K^-$ contour represents the $1\sigma$ errors in \eqref{Adir-1} and \eqref{Amix-1}.}
\label{fig:optimal}
\end{figure}

The analyses discussed in the previous two sections provide target regions for 
improved measurements at the Tevatron and for the early data taking at LHCb. 
In the long run, the major application of the $B^0_s\to K^+K^-$ decay will be
the determination of $\gamma$. The key improvement with respect to the 
discussion in Section~\ref{sec:gam} will be made possible by the measurement 
of both $\Adir(\BsKK)$ and $\Amix(\BsKK)$ \cite{RF-BsKK} (the expected 
LHCb statistical sensitivity for $\Adir(\BsKK)$ is 0.15 for already 
$200\,\mbox{pb}^{-1}$ \cite{Wilkinson}). We can then convert the corresponding 
values into a contour in the $\gamma$--$d'$ plane that is theoretically clean, 
in analogy to the $\gamma$--$d$ contour following from the CP-violating observables 
of the $B^0_d\to \pi^+\pi^-$ channel shown in Fig.~\ref{fig:1}. Using then the $U$-spin 
relation $d'=d$, we can determine $\gamma$ and $d$ from the intersection of the contours, 
as well as $\theta$ and $\theta'$, allowing an internal consistency check of the 
$U$-spin assumption. In particular, the quantity $K$ does not enter this determination
of $\gamma$. 
It can then instead be used to extract $|{\cal C}'/{\cal C}|$ for comparison with
theoretical analyses of this ratio.

In Fig.~\ref{fig:optimal}, we illustrate the corresponding contour in the $\gamma$--$d'$
plane that is determined through the CP asymmetries of the $B^0_s\to K^+K^-$ channel.
To this end, we refer to the $\gamma$ analysis of Section~\ref{sec:gam}. Using the values 
of $\gamma$, $d$ and $\theta$ in (\ref{gam-extr}) and (\ref{eqn:gam-hadr}), and neglecting the corresponding $U$-spin-breaking errors, we obtain
\begin{eqnarray}
\Adir(\BsKK)&=&0.094^{+0.044}_{-0.039}, \label{Adir-1}\\
\Amix(\BsKK)|_{\rm SM}&=&-0.218^{+0.037}_{-0.036}. \label{Amix-1}
\end{eqnarray}
These numbers are fully consistent with (\ref{ABsKK-input}) and (\ref{Amix-SM}), which 
rely on different inputs, thereby further supporting our numerical analysis. The band
in Fig.~\ref{fig:optimal} referring to the CP-violating $B^0_s\to K^+K^-$ observables
corresponds to the central values in (\ref{Adir-1}) and (\ref{Amix-1}) and their $1\sigma$
ranges. The $B^0_d\to\pi^+\pi^-$ contour is the same as in the left panel of Fig.~\ref{fig:1},
and, in order to guide the eye, we have also included the central value of the contour
fixed through $K$ and ${\cal A}_{\rm CP}^{\rm mix}(B_d\to\pi^+\pi^-)$. It is interesting to
observe that the $B^0_s\to K^+K^-$ and $B^0_d\to\pi^+\pi^-$ contours are intersecting with
a large angle, thereby leading again to a situation that is very robust with respect to 
possible $U$-spin-breaking corrections to $d'=d$. It will be interesting to confront the
contours in Fig.~\ref{fig:optimal} with future LHCb data.

\section{Conclusions}\label{sec:concl}
We have performed an analysis of the $U$-spin-related $B^0_s\to K^+K^-$, $B^0_d\to\pi^+\pi^-$
system in view of updated experimental and theoretical information and the first 
measurements from the LHCb experiment that are expected to arrive soon. 
We obtain a value of $\gamma=(68.3^{+4.9}_{-5.7}|_{\rm input}\mbox{}^{+5.0}_{-3.7}|_\xi
\mbox{}^{+0.1}_{-0.2}|_{\Delta\theta})^\circ$, which is very competitive with other direct 
determinations of this angle. Moreover, our result is in excellent agreement with the fits for the 
unitarity triangle, thereby excluding large NP effects at the decay amplitude level. 

A particularly interesting first observable of the $B^0_s\to K^+K^-$ decay to be measured 
at LHCb will be its effective lifetime $\tau_{K^+K^-}$. We have calculated this observable as 
a function of the $B^0_s$--$\bar B^0_s$ mixing phase and arrive at a picture that is remarkably 
stable both with respect to the current errors of the relevant input quantities and with respect
to possible $U$-spin-breaking effects. Moreover, we have shown that $\tau_{K^+K^-}$
offers an interesting alternative tool to get evidence for a sizeable NP value of $\phi_s$, 
thereby complementing the analyses of CP violation in $B^0_s\to J/\psi \phi$. 

The next newly measured observable to enter the stage should be the mixing-induced
CP asymmetry of $B^0_s\to K^+K^-$, which is correlated with $\sin\phi_s$ in an
interesting way. This correlation is again remarkably stable with respect to the errors of
the input quantities and possible $U$-spin-breaking corrections. Even for moderate values 
of $\phi_s$, $\Amix(\BsKK)$ deviates significantly from its SM value and therefore provides 
another sensitive probe for CP-violating NP effects in $B^0_s$--$\bar B^0_s$ mixing. The 
measurement of $\Amix(\BsKK)$ will then also allow us to determine $\phi_s$ unambiguously 
and, in particular, to distinguish the SM case with $\phi_s\sim 0^\circ$ from a NP scenario 
with $\phi_s\sim180^\circ$, both leading to small CP violation in $B^0_s\to J/\psi \phi$. 

Finally, once the direct CP asymmetry of $B^0_s\to K^+K^-$ is also accurately measured, we can optimize the extraction of the angle $\gamma$ of the unitarity triangle, which will be
the major application of the $B^0_s\to K^+K^-$, $B^0_d\to\pi^+\pi^-$ system at LHCb in
the long run. Moreover, we can then also perform internal checks of the $U$-spin symmetry
assumption. The picture following from the current data points towards a stable and
favourable situation with respect to possible $U$-spin-breaking effects. 

We look forward to confronting the ``target regions" in observable space calculated in 
this paper with real experimental data!

\vspace*{0.5truecm}

\noindent
{\it Acknowledgements}\\
We would like to thank Marco Gersabeck, Tristan du Pree and Gerhard Raven for 
useful discussions.

\end{document}